\documentclass[12pt]{article}

\usepackage{feynmp}                       

\usepackage[dvips]{color}                 
\usepackage{graphicx}                     
\usepackage{amssymb}

\usepackage{xspace}                       

\newcommand{\ia}{{\"{\i}}}   
\newcommand{\absatz}{\vspace{2ex}\noindent}

\renewcommand{\today}{\ifcase\day\or 1st\or 2nd\or 3rd\or 4th\or 5th\or 6th\or
        7th\or 8th\or 9th\or 10th\or 11th\or 12th\or 13th\or 14th\or 15th\or 
        16th\or 17th\or 18th\or 19th\or 20th\or 21st\or 22nd\or 23rd\or 24th\or
        25th\or 26th\or 27th\or 28th\or 29th\or 30th\or 
        31st\fi~\ifcase\month\or January\or February\or March\or April\or
        May\or June\or July\or August\or September\or October\or November\or
        December\fi \space \number\year}   

\newcommand{\mybibsing}[2]{#1: #2.}
\newcommand{\mybibmult}[3]{#1, #2: #3.} 
\newcommand{\myjournal}[2]{#1 [#2]}
\newcommand{\journal}[4]{{#1}\textbf{#2}, #3 (#4)}
\newcommand{\FBSS}{\emph{Few-Body Syst.\ }\textbf{Suppl.}}

\newcommand{\NPA}{\emph{Nucl.\ Phys.\ }\textbf{A}}
\newcommand{\NPB}{\emph{Nucl.\ Phys.\ }\textbf{B}}
\newcommand{\PLB}{\emph{Phys.\ Lett.\ }\textbf{B}}
\newcommand{\PR}{\emph{Phys.\ Rev.\ }}

\newcommand{\PRC}{\PR\textbf{C}}
\newcommand{\PRD}{\PR\textbf{D}}
\newcommand{\PRL}{\PR\emph{Lett.\ }}

\newcommand{\dis}{\displaystyle}

\newcommand{\ii}{\mathrm{i}}

\newcommand{\kv}{\vec{k}}

\newcommand{\mpi}{m_\pi}
\newcommand{\fpi}{f_\pi}
\newcommand{\MeV}{\mathrm{MeV}}
\newcommand{\fm}{\mathrm{fm}}

\newcommand{\NtwoLO}{N${}^2$LO\xspace}

\newcommand{\Qorder}{Q/\Lambda}

\newcommand{\calA}{\mathcal{A}} \newcommand{\calL}{\mathcal{L}}
 \newcommand{\calO}{\mathcal{O}}

\newcommand{\mytitle}[1]{
                         \begin{center}
                           \LARGE{\textbf{#1}}
                         \end{center}}
\newcommand{\myauthor}[1]{\textbf{#1}}
\newcommand{\myaddress}[1]{\textit{#1}}
\newcommand{\mypreprint}[1]{\begin{flushright} #1 \end{flushright}}

\begin{document}

\begin{fmffile}{comfeyn}
  \fmfset{curly_len}{2mm} \fmfset{dash_len}{1.5mm} \fmfset{wiggly_len}{3mm}
  \newcommand{\feynbox}[2]{\mbox{\parbox{#1}{#2}}}
  \newcommand{\fs}{\scriptstyle}
  \newcommand{\hq}{\hspace{0.5em}} \newcommand{\hqm}{\hspace{-0.25em}}
  
  \fmfcmd{vardef ellipseraw (expr p, ang) = save radx; numeric radx; radx=6/10
    length p; save rady; numeric rady; rady=3/10 length p; pair center;
    center:=point 1/2 length(p) of p; save t; transform t; t:=identity xscaled
    (2*radx*h) yscaled (2*rady*h) rotated (ang + angle direction length(p)/2
    of p) shifted center; fullcircle transformed t enddef;
    style_def ellipse expr p= shadedraw ellipseraw (p,0); enddef;}
  
%

\begin{titlepage}
  
  \mypreprint{
    NT@UW-00-022 \\
    TRI-PP-00-62\\
    TUM-T39-00-18 \\
    21st December 2000 }

  \mytitle{Nucleon Polarisabilities from \\
    Compton Scattering on the Deuteron}
  
  \vspace*{0.5cm}

\begin{center}
  
  \myauthor{Harald W.\ Grie\3hammer}~${}^{a,\,}$\footnote{Email:
    hgrie@physik.tu-muenchen.de} and \myauthor{Gautam
    Rupak}~${}^{b,\,c,\,}$\footnote{Email: grupak@triumf.ca; permanent
    address: (c)}
  
  \vspace*{0.5cm}
  
  \myaddress{
    ${}^a$Institut f{\"u}r Theoretische Physik, Physik-Department,\\
    Technische Universit{\"a}t M{\"u}nchen, D-85747 Garching, Germany\\
    ${}^b$Nuclear Theory Group,
    Department of Physics, University of Washington,\\
    Box 351560, Seattle, WA 98195-1560, USA\\
    ${}^c$TRIUMF, Vancouver, B.C., V6T 2A3, Canada}
  
  \vspace*{0.2cm}

\end{center}

\vspace*{0.5cm}

\begin{abstract}
  An analytic calculation of the differential cross section for elastic
  Compton scattering on the deuteron at photon energies $\omega$ in the range
  of $25-50\;\MeV$ is presented to next-to-next-to-leading order, i.e.~to an
  accuracy of $\sim 3\%$.  The calculation is model-independent and performed
  in the low energy nuclear Effective Field Theory without dynamical pions.
  The iso-scalar, scalar electric and magnetic nucleon polarisabilities
  $\alpha_0$ and $\beta_0$ enter as free parameters with a theoretical
  uncertainty of about $20\%$. Using data at $\omega_\mathrm{Lab}=49\;\MeV$ we
  find $\alpha_0=8.4\pm 3.0(\mathrm{exp})\pm 1.7(\mathrm{theor})$,
  $\beta_0=8.9\pm 3.9(\mathrm{exp})\pm 1.8(\mathrm{theor})$, each in units of
  $10^{-4}\;\fm^3$. With the experimental constraint for the iso-scalar Baldin
  sum rule, $\alpha_0=7.2\pm 2.1(\mathrm{exp})\pm 1.6(\mathrm{theor})$,
  $\beta_0=6.9\mp 2.1(\mathrm{exp})\mp 1.6(\mathrm{theor})$. A more accurate
  result can be achieved by: (i) better experimental data, and (ii) a higher
  order theoretical calculation including contributions from a couple of so
  far undetermined four-nucleon-two-photon operators.
\end{abstract}
\vskip 1.0cm
\noindent
\begin{tabular}{rl}
Suggested PACS numbers:& 13.60.Fz, 14.20.Dh, 21.30.Fe, 25.20.-x\\[1ex]
Suggested Keywords: &\begin{minipage}[t]{11cm}
                    Effective Field Theory, nucleon and neutron
                    polarisability, deuteron Compton scattering
                    \end{minipage}
\end{tabular}

\vskip 1.0cm

\end{titlepage}

\setcounter{page}{2} \setcounter{footnote}{0} \newpage
  
%


\absatz Albeit scalar electric and magnetic polarisabilities are fundamental
properties of the neutron, there is a fair amount of uncertainty in their
values for the lack of free neutron targets. Thus, one has to resort to
indirect methods to extract neutron polarisabilities. For example, Rose
et.~al. depend on a particular treatment of the neutron as a quasi-free
particle inside the deuteron and find for inelastic Compton scattering on the
deuteron $\gamma d\rightarrow\gamma n p$ that\footnote{In this letter, we
  express polarisability values in units of $10^{-4}\;\fm^3$.}
$\alpha^{(n)}=10.7^{+3.3}_{-10.7}$~\cite{Rose90b}.  Since the strong
electromagnetic field of a heavy nucleus could enhance small effects from
polarisability contributions, neutron scattering off a heavy nucleus is
another method employed.  However, analyses of low energy neutron-lead
scattering experiments give conflicting values: $\alpha^{(n)}=12.3\pm 1.5\pm
2.0$, $\beta^{(n)}=3.1\mp1.5\mp2.0$ in Ref.~\cite{SRHH91}, and more recently
as part of an investigation of additional nuclei $\alpha^{(n)}=0\pm 5$ in
Ref.~\cite{Koester95}.  In contradistinction, the scalar proton
polarisabilities are well-measured from Compton scattering off the
proton~\cite{Gibbon95}: $\alpha^{(p)}=12.1\pm 0.8\pm 0.5$ and
$\beta^{(p)}=2.1\mp 0.8\mp 0.5$, respectively.


On the theoretical side, a self-consistent and controlled description of the
polarisabilities is only encountered with the dawn of Chiral Perturbation
Theory ($\chi$PT)~\cite{BKM91}.  Unfortunately, estimates of corrections
beyond leading order (LO) in $\chi$PT are difficult due to unknown parameters
which already enter at next-to-leading order
(NLO)~\cite{Bernard:1993bg,Hemmert:1998tj}.  In this paper, we propose low
energy Compton deuteron scattering as a probe for neutron polarisabilities.
This is not a new idea and has fervently been advocated by Levchuk and L'vov,
by Weyrauch, and by Arenh\"ovel, Wilhelm and Wilbois, most recently in
Refs.~\cite{LevchukLvov,Weyrauch,AWW}; for its history see Ref.~\cite{KM99}
and references therein. That the $\chi$PT values for the iso-scalar
polarisabilities are consistent with Compton scattering data on the deuteron
was demonstrated in two Effective Field Theory (EFT) NLO calculations in
extensions of $\chi$PT to the few nucleon system, one using perturbative
pions~\cite{CGSSpol}, one using non-perturbative pions~\cite{Beaneetal}.

The present calculation is unique in the sense that all the theoretical
approximations are explicit and controlled, up to the order of the 
perturbative EFT calculation.  The polarisabilities extracted from this
calculation are independent of any modelling of the short-distance dynamics of
nucleons inside a deuteron or any heavy nucleus. In particular, they are
independent of assumptions about the pion-nucleon interaction, independent of
the form of the inter-nucleon potential, manifestly independent of the
photon coupling to nucleons and pions as extended objects and independent of
the choice of regulator.
The calculation is analytic and straightforward. 
To the accuracy claimed, the only experimental inputs are simple low energy
observables: the binding energy and pole residue of the deuteron, the nucleon
iso-vector magnetic moment and the nucleon-nucleon scattering length in the
${}^1\mathrm{S}_0$ channel. Our result will therefore -- to the accuracy
claimed -- be reproduced by any ``realistic'' potential model, and can hence
serve as cross check. However, at higher orders beyond those presented here,
there are contributions from certain four-nucleon-two-photon operators
representing short-distance physics that might not be reproduced accurately in
a model calculation. In conclusion, the polarisabilities are obtained in a
model independent way by a fit to data, and no explanation as to their origin
is attempted.

The letter is organised as follows: We start by outlining the framework of the
calculation and the power counting needed to determine the theoretical
accuracy. Then, the amplitude and cross section is presented. Finally, various
fits of the polarisabilities to data are discussed. A more detailed
presentation will be given in an upcoming article~\cite{GRSb}.



\absatz Our analysis is in the context of the few-nucleon EFT in which
nucleons and photons are the only dynamical degrees of freedom~\cite{CRSa}.
This formulation can be viewed as a systematisation of Effective Range Theory
to include interactions with gauge fields, and relativistic and short-distance
effects. Contributions from pions and other, heavier degrees of freedom are
included at low energy through multiple-nucleon-photon contact interactions in
perturbation.  Since one cannot apply this EFT at energies where cuts in the
amplitude are probed due to pion propagation, meson exchange etc., the natural
breakdown scale $\Lambda$ is the pion mass $m_\pi\approx 140\;\MeV$. For
details of the formalism, we refer to a recent review~\cite{review}.

EFTs live from the fact that at low energies, a separation of scales exists:
Like the binding momentum of the deuteron $\gamma= 45.7066\;\MeV$, the photon
momentum $\kv$ and energy $\omega= |\kv|$ are much smaller than $\Lambda$.
There is another ``high-energy'' scale, namely the iso-spin averaged nucleon
mass $M_N=938.92\;\MeV$. Thus, each physical low energy observable at a
typical momentum scale $Q\sim\gamma$ is expanded systematically in powers of
two small parameters: $\frac{Q}{\Lambda}$ and $\frac{Q}{M_N}$.  For
convenience, we formally identify $\Lambda/M_N\sim \Qorder$, which is
accidentally valid. Thus, relativistic corrections with a typical size
$\gamma^2/M_N^2\sim (Q/M_N)^2=(\Qorder)^2\times (\Lambda/M_N)^2$ contribute at
$\calO((\Qorder)^4)$, i.e.~N$^4$LO, see Refs.~\cite{CRSa,BG99,rupak99}.

The centre-of-mass photon energy $\omega$ and the conversion of relativistic
photon energies to non-relativistic nucleon energies introduce new scales:
$\omega$ and $\sqrt{\omega M_N}$.
Therefore, it is convenient to separate the calculation into two different
energy r\'egimes~\cite{CGSSpol}. In \emph{r\'egime I}, $\omega\lesssim
\gamma^2/M_N$, and $\sqrt{M_N\omega}\sim\gamma\sim Q$. We concentrate on
\emph{r\'egime II}, i.e.~higher photon energies $\omega\sim\gamma\sim Q$, in
which factors of $\sqrt{\omega M_N}\sim \Lambda$ are summed to all orders.  In
addition, we make some integral approximations that allow us to effortlessly
move from one r\'egime to another with one simple expression~\cite{CGSSpol}.
Details of the power counting, calculation and validity of the approximations
will be provided in a future publication~\cite{GRSb}.

In the Weyl gauge $A_0=0$, the amplitude for Compton scattering can be written
as
\begin{eqnarray}
  \label{ComptonAmplitude}
  \ii\calA&=& \ii \;\frac{e^2}{2M_N}\left[S\;
    \vec{\varepsilon}_d\cdot\vec{\varepsilon}^{\prime\ast}_d
    +V_i\;\epsilon_{ijk}\;\varepsilon_{d,j}
    \varepsilon^{\prime\ast}_{d,k}
    +T_{ij}\;\left(\varepsilon_{d,i}\varepsilon^{\prime\ast}_{d,j}
      +\varepsilon_{d,j}\varepsilon^{\prime\ast}_{d,i}
      -\frac{2}{3}\delta_{ij}\ 
      \vec{\varepsilon}_{d}\cdot\vec{\varepsilon}^{\prime\ast}_{d}\right)
  \right]\;\;,
\end{eqnarray}
representing the scalar $S$, vector $V$ and tensor $T$ coupling of a photon
with incoming (outgoing) polarisation $\vec{\varepsilon}_\gamma$
($\vec{\varepsilon}_\gamma^\prime$) with a deuteron of incoming (outgoing)
polarisation $\vec{\varepsilon}_d$ ($\vec{\varepsilon}_d^\prime$).
$\frac{e^2}{4\pi}=\alpha=1/137$. The power counting reveals that the vector
and tensor amplitudes start contributing to the unpolarised cross section only
at N${}^4$LO. Hence the most dominant term is the scalar one. The
centre-of-mass scalar amplitude $S$ is conventionally written as
\begin{eqnarray}
  \label{ComptonScalar}
  S&=& F(\omega, \theta)\;
  \vec{\varepsilon}_\gamma\cdot\vec{\varepsilon}^{\prime\ast}_\gamma +
  G(\omega,\theta)\;\left(\hat{\kv}\times\vec{\varepsilon}_\gamma\right)\cdot
  \left(\hat{\kv^\prime}\times\vec{\varepsilon}^{\prime\ast}_{\gamma}\right)
  \;\;,
\end{eqnarray}
where $F$ and $G$ represent the scalar electric and magnetic amplitudes, and
$\theta$ is the angle between the directions of the incoming and outgoing
photon momenta, $\hat{\kv}$ and $\hat{\kv^\prime}$.  The familiar Thomson
limit is reproduced by $F(\omega=0)=1$ and $G(\omega=0)=0$.

For $F$ and $G$, an expansion in powers of $\Qorder$ exists,
$F=F^{(0)}+F^{(1)}+\dots$ etc., where the superscript denotes the contribution
of order $(\Qorder)^n$.  We keep all N$^2$LO, i.e.~$(\Qorder)^2$, corrections
compared to LO, $\calO((\Qorder)^0)$.  One such effect is the deuteron wave
function renormalisation, introducing factors of $Z_d-1$. $Z_d=1.690(3)$ is
the residue of the scattering amplitude at the deuteron pole and corresponds
to the normalisation of the deuteron wave function at large wave length.
$Z_d-1$ is treated as $\calO(\Qorder)$, see Ref.~\cite{PRS} for details.
Higher order corrections such as relativistic effects and
${}^3\mathrm{S}_1$-${}^3\mathrm{D}_1$ mixing are irrelevant for the present
N$^2$LO calculation~\cite{CRSa,rupak99}. Assuming that all contributions not
taken into account are of natural size, we expect our result to be accurate to
order $(\Qorder)^3\lesssim 0.03$.  This error estimate lies at the heart of a
model-independent, controlled approximation provided by EFT.


\absatz Now for details of the calculation, performed in the centre-of-mass
frame. The contribution from the seagull diagram in Fig.~\ref{diagrams.fig}
(a) scales as $\calO((\Qorder)^0)$, which constitutes a LO contribution in
both energy r\'egimes. We get
\begin{eqnarray}
  &&F^{(0)}_a= \frac{4\,{\sqrt{2}}\,\gamma \,
  }{\omega \, {\sqrt{1 - \cos\theta}}}\;\arctan \left[\frac{\omega \,
    {\sqrt{1- \cos\theta}}}{2\,
  {\sqrt{2}}\,\gamma }\right]\;\;, \nonumber\\[1.5ex]
\label{ComptonLO.eq}
  &&F^{(1)}_a= (Z_d-1) F^{(0)}_a\;\;,\;\;F^{(2)}_a=0\;\;,\;\;
  G_a=0\;\; .
\end{eqnarray}
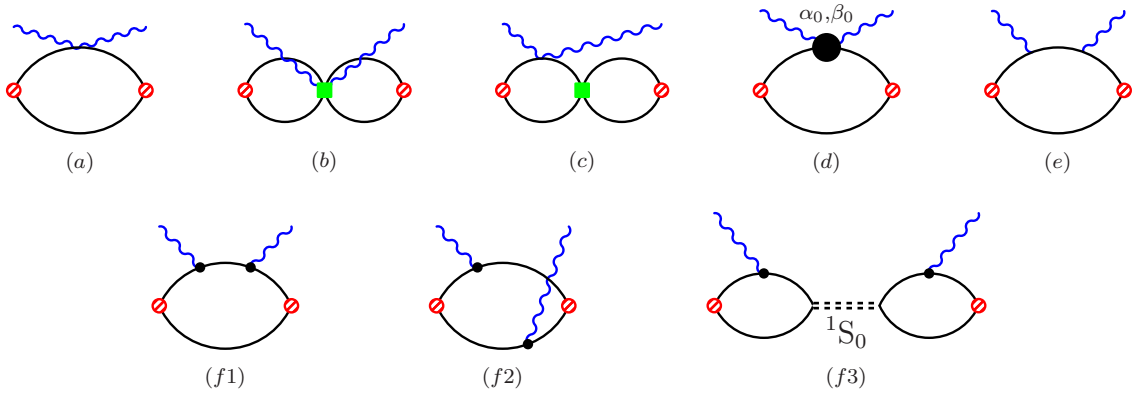
\begin{figure}[!htb]
  \begin{center}
    \begin{minipage}{0.9\linewidth}
      \feynbox{50\unitlength}{
            \begin{fmfgraph*}(50,60)
              \fmfleft{i} \fmfright{o} \fmftop{t1,t2}
              \fmf{vanilla,width=thin,left=0.65}{i,o}
              \fmf{vanilla,width=thin,left=0.65,label=$\fs(a)$,label.side=left,
                label.dist=7}{o,i} \fmffreeze
              \fmfv{decor.shape=circle,decor.filled=shaded,foreground=red,
                decor.size=5}{i}
              \fmfv{decor.shape=circle,decor.filled=shaded,foreground=red,
                decor.size=5}{o} \fmffreeze \fmfipath{pa}
              \fmfiset{pa}{vpath(__i,__o)} \fmffreeze
              \fmfi{photon,foreground=blue}{point 1/2 length(pa) of pa --
                vloc(__t1)} \fmfi{photon,foreground=blue}{point 1/2 length(pa)
                of pa -- vloc(__t2)}
              \end{fmfgraph*}}
            \hfill
            \feynbox{60\unitlength}{
            \begin{fmfgraph*}(60,50)
              \fmfstraight \fmfleft{i1,i2,i3,i4,i5} \fmfright{o1,o2,o3,o4,o5}
              \fmf{vanilla,width=thin,left=0.8}{v,i3}
              \fmf{vanilla,width=thin,left=0.8}{i3,v}
              \fmf{vanilla,width=thin,left=0.8}{o3,v}
              \fmf{vanilla,width=thin,left=0.8}{v,o3} \fmffreeze
              \fmfv{decor.shape=square,decor.size=5,foreground=green,
                label=${\fs (b)}$,label.angle=-90,label.dist=23}{v}
              \fmfv{decor.shape=circle,decor.filled=shaded,foreground=red,
                decor.size=5}{i3}
              \fmfv{decor.shape=circle,decor.filled=shaded,foreground=red,
                decor.size=5}{o3} \fmffreeze
              \fmf{photon,foreground=blue}{i5,v,o5}
            \end{fmfgraph*}}
          \hfill
          \feynbox{60\unitlength}{
            \begin{fmfgraph*}(60,50)
              \fmfstraight \fmfleft{i1,i2,i3,i4,i5} \fmfright{o1,o2,o3,o4,o5}
              \fmf{vanilla,width=thin,left=0.8}{v,i3}
              \fmf{vanilla,width=thin,left=0.8}{i3,v}
              \fmf{vanilla,width=thin,left=0.8}{o3,v}
              \fmf{vanilla,width=thin,left=0.8}{v,o3} \fmffreeze
              \fmfv{decor.shape=square,decor.size=5,foreground=green,
                label=${\fs (c)}$,label.angle=-90,label.dist=23}{v}
              \fmfv{decor.shape=circle,decor.filled=shaded,foreground=red,
                decor.size=5}{i3}
              \fmfv{decor.shape=circle,decor.filled=shaded,foreground=red,
                decor.size=5}{o3} \fmffreeze \fmfipath{pa} \fmfipath{pb}
              \fmfiset{pa}{vpath(__i3,__v)} \fmfiset{pb}{vpath(__v,__i3)}
              \fmfi{photon,foreground=blue}{point 1/2 length(pa) of pa --
                vloc(__i5)} \fmfi{photon,foreground=blue}{vloc(__o5) -- point
                1/2 length(pa) of pa}
            \end{fmfgraph*}}        
          \hfill
          \feynbox{50\unitlength}{
            \begin{fmfgraph*}(50,60)
              \fmfleft{i} \fmfright{o} \fmftop{t1,t2}
              \fmf{vanilla,width=thin,left=0.65}{i,o}
              \fmf{vanilla,width=thin,left=0.65,label=$\fs(d)$,label.side=left,
                label.dist=7}{o,i} \fmffreeze
              \fmfv{decor.shape=circle,decor.filled=shaded,foreground=red,
                decor.size=5}{i}
              \fmfv{decor.shape=circle,decor.filled=shaded,foreground=red,
                decor.size=5}{o} \fmffreeze \fmfipath{pa}
              \fmfiset{pa}{vpath(__i,__o)} \fmfipair{upper}
              \fmfiset{upper}{point 1/2 length(pa) of pa}
              \fmfi{photon,foreground=blue}{upper -- vloc(__t1)}
              \fmfi{photon,foreground=blue}{upper -- vloc(__t2)}
              \fmfiv{decor.shape=circle,decor.filled=full,decor.size=10,
                label=$\fs\alpha_0,,\beta_0$,label.angle=90,label.dist=9
                }{upper}
              \end{fmfgraph*}}
            \hfill
            \feynbox{50\unitlength}{
            \begin{fmfgraph*}(50,60)
              \fmfleft{i} \fmfright{o} \fmftop{t1,t2}
              \fmf{vanilla,width=thin,left=0.65}{i,o}
              \fmf{vanilla,width=thin,left=0.65,label=$\fs(e)$,label.side=left,
                label.dist=7}{o,i} \fmffreeze
              \fmfv{decor.shape=circle,decor.filled=shaded,foreground=red,
                decor.size=5}{i}
              \fmfv{decor.shape=circle,decor.filled=shaded,foreground=red,
                decor.size=5}{o} \fmffreeze \fmfipath{pa}
              \fmfiset{pa}{vpath(__i,__o)} \fmffreeze
              \fmfi{photon,foreground=blue}{point 1/3 length(pa) of pa --
                vloc(__t1)} \fmfi{photon,foreground=blue}{point 2/3 length(pa)
                of pa -- vloc(__t2)}
              \end{fmfgraph*}}
            \\[3ex]
            \hspace*{\fill}
            \feynbox{50\unitlength}{
            \begin{fmfgraph*}(50,60)
              \fmfleft{i} \fmfright{o} \fmftop{t1,t2}
              \fmf{vanilla,width=thin,left=0.65}{i,o}
              \fmf{vanilla,width=thin,left=0.65,label=$\fs(f1)$,
                label.side=left,label.dist=7}{o,i} \fmffreeze
              \fmfv{decor.shape=circle,decor.filled=shaded,foreground=red,
                decor.size=5}{i}
              \fmfv{decor.shape=circle,decor.filled=shaded,foreground=red,
                decor.size=5}{o} \fmffreeze \fmfipath{pa}
              \fmfiset{pa}{vpath(__i,__o)} \fmffreeze \fmfipair{uppera}
              \fmfiset{uppera}{point 1/3 length(pa) of pa} \fmfipair{upperb}
              \fmfiset{upperb}{point 2/3 length(pa) of pa}
              \fmfi{photon,foreground=blue}{uppera -- vloc(__t1)}
              \fmfi{photon,foreground=blue}{upperb -- vloc(__t2)}
              \fmfiv{decor.shape=circle,decor.filled=full,
                decor.size=3
                }{uppera}
              \fmfiv{decor.shape=circle,decor.filled=full,
                decor.size=3}{upperb}
              \end{fmfgraph*}}
            \hfill \feynbox{50\unitlength}{
            \begin{fmfgraph*}(50,60)
              \fmfleft{i} \fmfright{o} \fmftop{t1,t2}
              \fmf{vanilla,width=thin,left=0.65}{i,o}
              \fmf{vanilla,width=thin,left=0.65,label=$\fs(f2)$,
                label.side=left,label.dist=7}{o,i} \fmffreeze
              \fmfv{decor.shape=circle,decor.filled=shaded,foreground=red,
                decor.size=5}{i}
              \fmfv{decor.shape=circle,decor.filled=shaded,foreground=red,
                decor.size=5}{o} \fmffreeze \fmfipath{pa}
              \fmfiset{pa}{vpath(__i,__o)} \fmfipath{pb}
              \fmfiset{pb}{vpath(__o,__i)} \fmffreeze \fmfipair{uppera}
              \fmfiset{uppera}{point 1/3 length(pa) of pa} \fmfipair{lowera}
              \fmfiset{lowera}{point 1/3 length(pb) of pb}
              \fmfi{photon,foreground=blue}{uppera -- vloc(__t1)}
              \fmfi{photon,foreground=blue}{lowera -- vloc(__t2)}
              \fmfiv{decor.shape=circle,decor.filled=full,
                decor.size=3}{uppera}
              \fmfiv{decor.shape=circle,decor.filled=full,
                decor.size=3}{lowera}
              \end{fmfgraph*}}
            \hfill \feynbox{100\unitlength}{
            \begin{fmfgraph*}(100,70)
              \fmfleft{i} \fmfright{o} \fmftop{t1,t2}
              \fmf{vanilla,width=thin,left=0.65}{i,v1}
              \fmf{vanilla,width=thin,left=0.65}{v1,i}
              \fmf{dbl_dashes,width=thin,label=${\dis{}^1\mathrm{S}_0}\atop
                \rule{0ex}{12\unitlength}{\fs (f3)}$,
                label.side=right,label.dist=4,tension=3}{v1,v4}
              \fmf{vanilla,width=thin,left=0.65}{v4,o}
              \fmf{vanilla,width=thin,left=0.65}{o,v4} \fmffreeze
              \fmfv{decor.shape=circle,decor.filled=shaded,foreground=red,
                decor.size=5}{i}
              \fmfv{decor.shape=circle,decor.filled=shaded,foreground=red,
                decor.size=5}{o} \fmffreeze \fmfipath{pa}
              \fmfiset{pa}{vpath(__i,__v1)} \fmfipath{pb}
              \fmfiset{pb}{vpath(__v4,__o)} \fmffreeze \fmfipair{uppera}
              \fmfiset{uppera}{point 1/2 length(pa) of pa} \fmfipair{lowera}
              \fmfiset{lowera}{point 1/2 length(pb) of pb}
              \fmfi{photon,foreground=blue}{uppera -- vloc(__t1)}
              \fmfi{photon,foreground=blue}{lowera -- vloc(__t2)}
              \fmfiv{decor.shape=circle,decor.filled=full,
                decor.size=3}{uppera}
              \fmfiv{decor.shape=circle,decor.filled=full,
                decor.size=3}{lowera}
              \end{fmfgraph*}}
            \hspace*{\fill}
\end{minipage}
            
\caption{\textit{Contributions to Compton scattering on the deuteron
    (represented by the circles) up to order \protect$(\Qorder)^2$.  Solid
    lines: nucleons; wavy lines: photons; squares: four-nucleon contact
    interactions needed to reproduce the correct deuteron residue; large disk:
    interactions via nucleon polarisabilities; dots: magnetic (Fermi)
    interactions; dashed double line: ${}^1\mathrm{S}_0$ di-baryon. Graphs
    with a permutation of external lines or vertices are not displayed.}}
\label{diagrams.fig}
  \end{center} 
  \vspace*{-4ex}
\end{figure}

Figures \ref{diagrams.fig} (b) and (c) contribute at NLO in both r\'egimes,
and up to \NtwoLO
\begin{eqnarray}
  \label{ComptonNLObc.eq}
  F_{b+c}=-(Z_d-1)&,&
  G_{b+c}=0\;\;.
\end{eqnarray}
Next, we consider the contributions from the iso-scalar, scalar electric and
magnetic nucleon polarisabilities with Lagrangean
$\calL_\mathrm{pol}=2\pi(\alpha_0 \vec{E}^2 +\beta_0\vec{B}^2)N^\dagger N$,
where $\alpha_0:=(\alpha^{(p)}+\alpha^{(n)})/2$ and
$\beta_0:=(\beta^{(p)}+\beta^{(n)})/2$. The polarisability diagrams in
Fig.~\ref{diagrams.fig} (d) contribute at different orders in the different
energy r\'egimes:
\begin{eqnarray}
  \label{Comptond.eq}
  F_d= -\frac{2 \alpha_0M_N}{\alpha} \;\omega^2\;F_a
  & ,&
  G_d= -\frac{2\beta_0M_N}{\alpha}\;\omega^2\; F_a\;\; 
\end{eqnarray}
The nucleon as point-like particle can only be polarised by its pion cloud,
and at LO in $\chi$PT $\alpha_0,\beta_0 \sim \alpha/(8\pi \fpi^2\mpi)$, i.e.
numerically $2\;(\alpha_0,\beta_0)\; M_N/\alpha\sim 1/\Lambda^2$ with
$\fpi=131\;\MeV\approx\mpi$~\cite{BKM91}. Thus, in r\'egime I these diagrams
are suppressed by $(\Qorder)^6$. In contradistinction, they contribute at
\NtwoLO, $(\Qorder)^2$, in r\'egime II.

The power counting for the diagrams in Fig.~\ref{diagrams.fig} (e) and (f)
is~slightly more involved because of factors of $\sqrt{\omega M_N}$ appearing
in the propagators of the loop integrals from the conversion of relativistic
photon energies to non-relativistic nucleon energies in the intermediate
nucleon propagators. Some diagrams contribute at different orders in the
different r\'egimes. For example, in r\'egime II ($\omega\sim\gamma$),
Fig.~\ref{diagrams.fig} (e) is of order $\gamma/\sqrt{\omega M_N}\approx
\Qorder$ and contributes at NLO to $F$.  Its contribution to the magnetic part
$G$ is suppressed by $(\Qorder)^2$, i.e.~\NtwoLO. But for small photon
energies $\omega\lesssim\gamma^2/M_N$, Siegert's theorem requires this diagram
to contribute as strongly as the seagull, Fig.~\ref{diagrams.fig} (a), because
of gauge invariance. Indeed, the contribution to $F$ from
Fig.~\ref{diagrams.fig} is $\calO((\Qorder)^0)$, i.e.~LO in r\'egime I; to $G$
it starts at N$^4$LO. We find for Fig.~\ref{diagrams.fig} (e) in r\'egime II
\begin{eqnarray}
  \label{ComptonNLOe.eq}
  F^{(LO)}_{e}&=& \frac{4 \gamma \left[2\ {{\gamma
      }^3}- {{({{\gamma }^2}-M_N \omega - \ii \epsilon)}^{3/2}}-
  {{( {{\gamma }^2}+M_N\omega )} ^{3/2}}\right]}{3M_N^2{{\omega }^2}}
  \;\;,\nonumber\\
  F^{(NLO)}_{e}&=& (Z_d-1)F^{(LO)}_{e}
  -\cos\theta\;G^{(LO)}_{e}\;\; ,\nonumber\\
  G^{(LO)}_{e}&=&\frac{1}{15M^4\,\omega^2}
  \left[8\gamma^6 + 20M\gamma^4\,\omega  + 60M^2\gamma^2\,\omega^2
    +15M^3\,\omega^3-4\gamma\left(\gamma^2-M\,\omega-\ii\epsilon\right)^{5/2}
    -\right.\nonumber\\
     &&\phantom{\frac{1}{30M^4\,\omega^2}} \;\;\left.-\gamma 
     \left(\gamma^2 + M\omega\right)^{1/2}\,
     \left(4\gamma^4+28M\gamma^2\,\omega+39M^2\,\omega^2 \right)\right]\ ,
\end{eqnarray}
where the approximations made for analytic results agree with the exact answer 
to within $0.5\%$.

The nuclear magnetic moment contributions from Fig.~\ref{diagrams.fig} (f) are
suppressed by factors of $\left(\kappa_1 Q/M_N\right)^2$ in both the energy
r\'egimes. Here, the square of the iso-vector nuclear magnetic moment in
nuclear magnetons, $\kappa_1^2=5.536$, is anomalously large, given an
expansion parameter of $\Qorder\approx 0.3$.
Numerically, Fig.~\ref{diagrams.fig} (f) is therefore expected to contribute
at \NtwoLO, and we choose to include it. The iso-scalar magnetic moment
$\kappa_0\approx 0.4$ does not enter at \NtwoLO because
it is of natural size.  Fig.~\ref{diagrams.fig} (f3) contains intermediate two
nucleon scattering through the ${}^1\mathrm{S}_0$ channel with a scattering
length of $a^{({}^1\mathrm{S}_0)}=-23.714\;\fm\sim 1/Q$. Analytic and exact
results for the momentum integrals agree again to better than $0.5\%$.
\begin{eqnarray}
  \label{ComptonNNLOf.eq}
  &&
  F_f^{(2)}=0\;\;,\nonumber\\
  &&
  G_f^{(2)}=-\kappa_1^2\;\frac{4\gamma}{3M_N^2}\;
  \left(\gamma-\frac{1}{a^{({}^1\mathrm{S}_0)}}\right)
  \left[\frac{\sqrt{\gamma^2+M_N\omega}-\gamma}{
  \frac{1}{a^{({}^1\mathrm{S}_0)}}-\sqrt{\gamma^2+M\omega}}
  +\frac{\sqrt{\gamma^2-M_N\omega- \ii \epsilon }-\gamma}{
    \frac{1}{a^{({}^1\mathrm{S}_0)}}-\sqrt{\gamma^2-M\omega- \ii \epsilon}}
  \right]\;\;\;\;
\end{eqnarray}
To summarise, in r\'egime II ($\omega\sim\gamma$) the electric and magnetic
amplitudes to \NtwoLO are
\begin{eqnarray}
  \label{regimeII.eq}
&&F^{(0)}= F^{(0)}_a\;\; ,\nonumber\\
&&F^{(1)}= F^{(1)}_a+F^{(1)}_{b+c}+F^{(LO)}_e\;\;,\nonumber\\
&&F^{(2)}= F^{(2)}_d+F^{(NLO)}_e+F^{(2)}_f\;\;,\nonumber\\
&&G^{(0)}=0\;\;,\;\;G^{(1)}=0\;\;,\;\;
     G^{(2)}= G^{(2)}_d+G^{(LO)}_e+G^{(2)}_f \;\;.
\end{eqnarray}



\absatz As noted above, the scalar polarisability contributions to Compton
scattering are suppressed by $(\omega/\Lambda)^2$. Up to \NtwoLO,
Compton scattering is therefore sensitive to the nucleon polarisabilities
$\alpha_0$, $\beta_0$ only in r\'egime II, where they enter as the only
undetermined parameters, see (\ref{Comptond.eq}) and (\ref{regimeII.eq}). Thus
photon energies $\omega\sim \gamma\sim 45\;\MeV$ are most appropriate for
extracting nucleon polarisabilities from a comparison of the differential
cross section in the centre-of-mass frame
\begin{eqnarray}
  \label{diffcross} 
  \frac{d\sigma}{d\Omega}\bigg|_\mathrm{cm}&=& \frac{\alpha^2}
  {2\left(\omega+\sqrt{\omega^2+4 M_N^2}\right)^2}
  \;\left[\left(|F|^2+|G|^2\right)(1+\cos^2\theta)
    +4\mathrm{Re}[FG^{\ast}]\cos\theta\right]\ ,
\end{eqnarray}
expanded to \NtwoLO, with experimental data.
Using the experimental values for the proton polarisabilities~\cite{Gibbon95},
we can then determine the neutron polarisabilities.

\begin{figure}[!htb]
  \begin{center}
    \parbox[c][\totalheight][t]{0.49\textwidth}{
      \includegraphics*[width=0.49\textwidth]{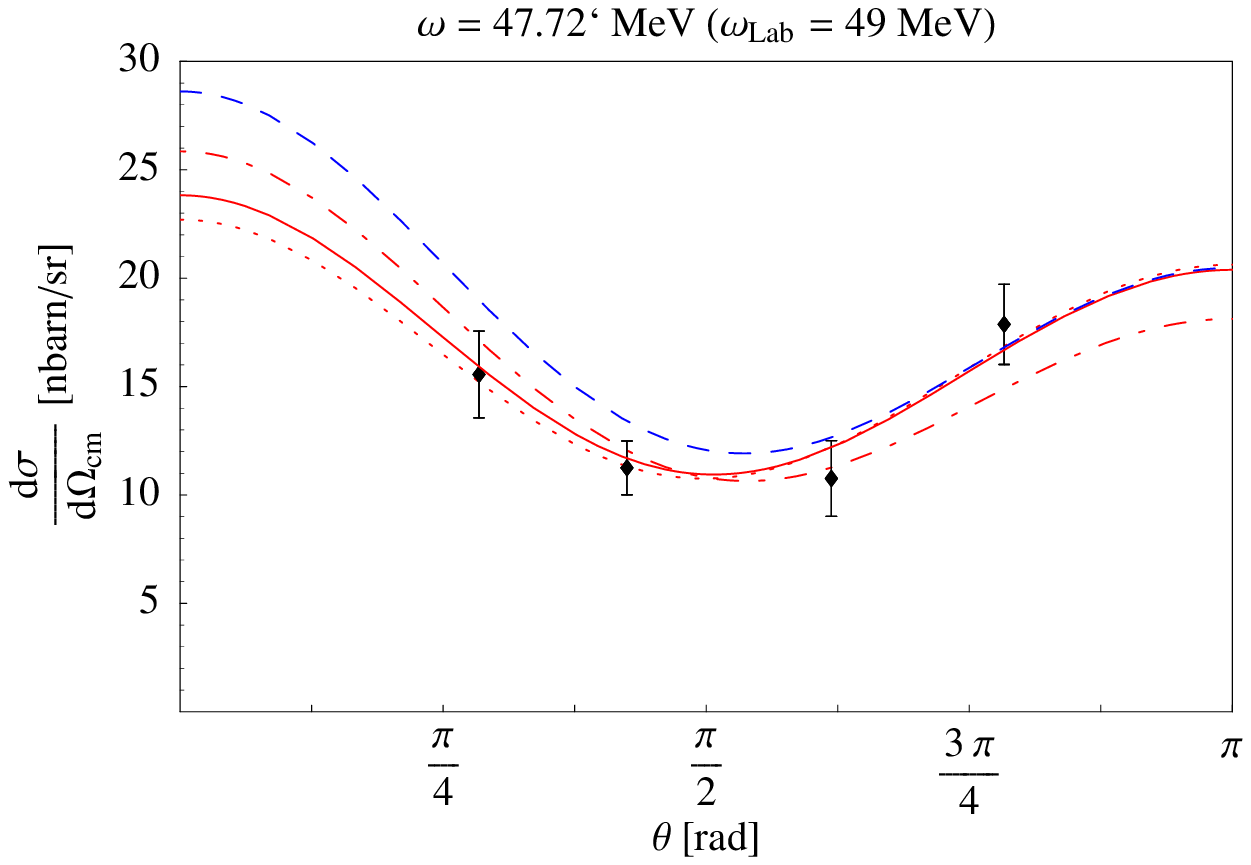} } \hfill
    \parbox[c][\totalheight][t]{0.49\textwidth}{
      \includegraphics*[width=0.49\textwidth]{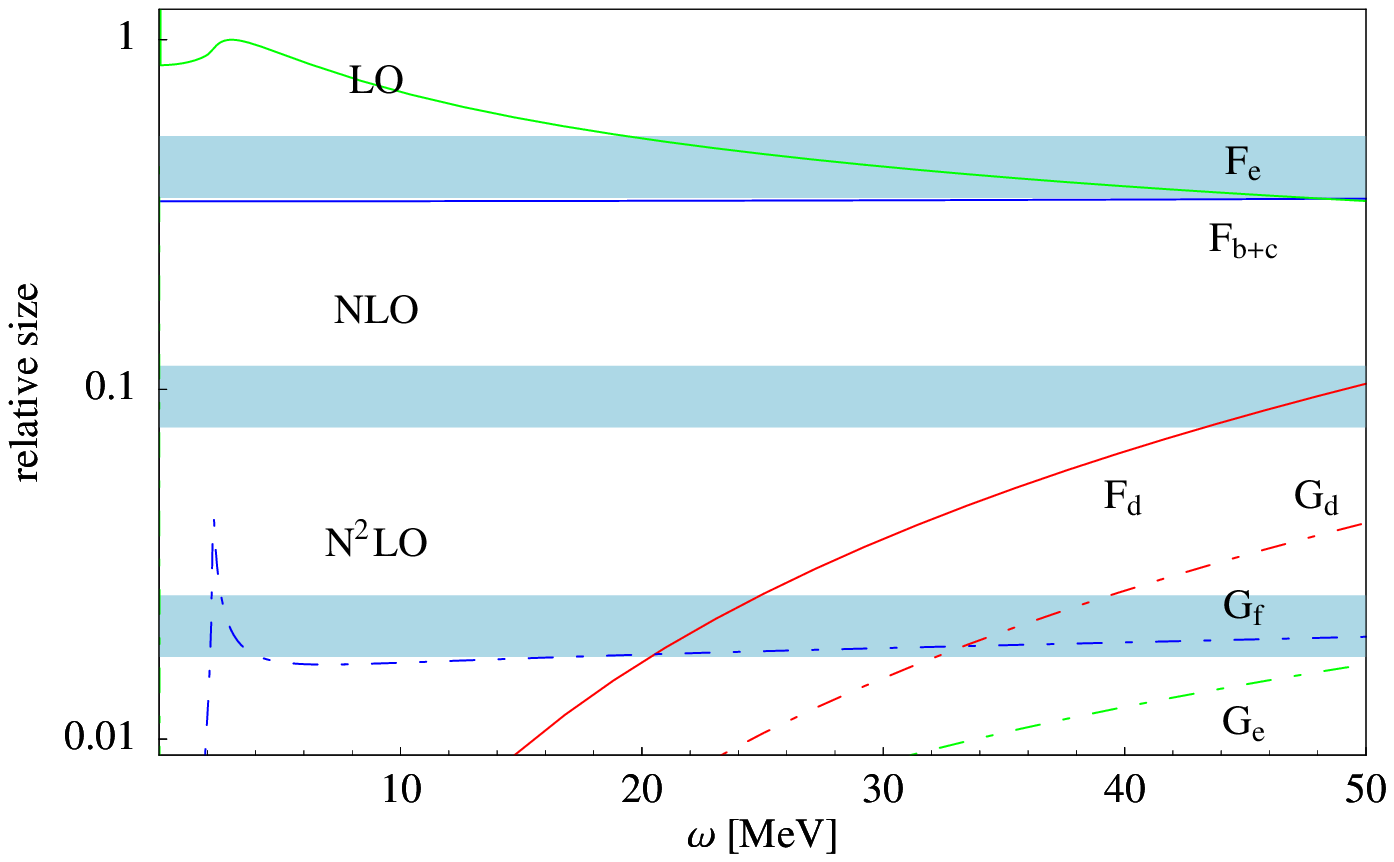} }
\caption{\textit{Left: EFT result for the differential cross section fitted
    to data~\cite{Lucas}. Dashed: \protect$\alpha_0=\beta_0=0$; dot-dashed:
    \protect$\beta_0=0$; dotted: \protect$\alpha_0$ and \protect$\beta_0$
    fitted; solid: \protect$\alpha_0$ and \protect$\beta_0$ constrained by the
    Baldin sum rule. Right: Sizes of the (angle averaged) contributions $F_i$
    and $G_i$, normalised to the LO contribution of the seagull term:
    $|F_i,G_i|/|F_a^{(0)}|$. The values assumed for the polarisabilities are
    $12$ and $5$. The grey bars are meant to guide the eye in determining the
    size of a contribution in the \protect$\Qorder$ expansion.}}
\label{comptonfit.fig}
  \end{center} 
  \vspace*{-4ex}
\end{figure}

We now proceed to determine the polarisabilities from Compton scattering data
at $\omega_\mathrm{Lab}=49\;\MeV$~\cite{Lucas}. If the LO $\chi$PT result
$\alpha_0=12.4,\; \beta_0=\frac{1}{10} \alpha_0$~\cite{BKM91} is taken at face
value, one might assume that $\beta\approx0$ at N$^2$LO. A $\chi^2$ fit gives
$\alpha_0=8.1\pm4.6$, with sizable $1\sigma$ errors\footnote{Unless otherwise
  stated, the errors are only from the fit and do not include the theoretical
  uncertainties.}. Estimates of the NLO corrections in $\chi$PT seem to favour
the electric and magnetic polarisabilities to have about the same size:
Resonance saturation suggests $\alpha_0=12.0\pm 2.5$ and
$\beta_0=5.7\pm5.1$~\cite{Bernard:1993bg}, and treating the $\Delta$ as
dynamical degree of freedom leads to $\alpha_0=16.4$,
$\beta_0=9.1$~\cite{Hemmert:1998tj}. In the next step, we therefore fit both
$\alpha_0$ and $\beta_0$ as free parameters, resulting in
$\alpha_0=8.4\pm3.0\pm1.7$ and $\beta_0=8.9\pm3.9\pm1.8$, again with large
error bars from a two parameter fit to four data points. The first error comes
from the fit, the second is the estimate of the theoretical uncertainties.
However, the well measured Baldin sum rule constrains
$\alpha_0+\beta_0=14.05\pm0.67$~\cite{Babusci:1998ij} which is also
comfortably close to the sum of the polarisabilities from the na{\ia}ve
two-parameter fit. Utilising the sum rule, we find $\alpha_0=7.2\pm2.1$ and
hence $\beta_0=6.9\mp2.1$ in a one parameter fit, with the best $\chi^2$ of
all the three procedures. Strictly speaking, the Baldin sum rule gives the sum
of the polarisabilities only at zero photon energy, while the polarisabilities
are extracted at $\omega=47.7\;\MeV$. However, we assume that corrections are
small~\cite{hgth}. Figure~\ref{comptonfit.fig} compares the three fits to
scattering data.

The calculation presented here is formally accurate to $\calO((\Qorder)^3)$,
i.e.~to within $\approx 3\%$. However, numerically the result might be
slightly better as some contributions with sizes comparable to N$^3$LO are
already included in (\ref{ComptonNLOe.eq}) and (\ref{ComptonNNLOf.eq}) and the
polarisabilities are by far the biggest \NtwoLO contributions.
Figure~\ref{comptonfit.fig} compares the strengths of the diagrams, confirming
the power counting and demonstrating convergence.  We can for example leave
out the $\kappa_1$ contribution Fig.~\ref{diagrams.fig} (f) which acts like a
large N${}^3$LO correction, see Fig.~\ref{comptonfit.fig}. This enhances the
electric polarisability using the Baldin sum rule to $\alpha_0=8.1\pm2.1$ and
reduces $\beta_0$ accordingly. On the other hand, one might expect that wave
function renormalisation effects on the polarisability diagrams are large
because $Z_d-1=0.69\dots$ is large for a NLO effect. Still, the fit using the
Baldin sum rule then induces only small changes in the polarisabilities,
$\alpha_0=7.3\pm2.1$.  We therefore estimate the theoretical uncertainty of
our extraction at $\omega\approx 40\;\MeV$ to be about $20\%$ and refer to
\cite{GRSb} for further discussion. The experimental uncertainty is at present
higher than the theoretical error because of the comparatively high error bars
in the only available low energy Compton scattering data.

Including the theoretical uncertainties in our calculation ($\approx 20\%$)
and in the value of the Baldin sum rule, we find $\alpha_0=7.2\pm2.1\pm 1.4\pm
0.7$, $\beta_0=6.9\mp 2.1\mp 1.4\mp 0.7$, where the first error comes from the
fit, the second one is the theoretical uncertainty of our \NtwoLO calculation,
and the third comes from applying the Baldin sum rule. For curiosity, we note
that these values are also well consistent with the two Urbana data points at
$\omega_\mathrm{Lab}=69\;\MeV$~\cite{Lucas} which give using the Baldin sum
rule $\alpha_0=5.4\pm2.2$, $\beta_0=8.6\mp2.2$. 
We finally extract for the neutron polarisabilities
$\alpha^{(n)}\approx(5;\,2)$ and $\beta^{(n)}\approx(16;\,12)$, where the
first number in parentheses comes from the free fit, and the second from the
fit using the Baldin sum rule. We quote no errors since they are large mainly
due to the large experimental uncertainties.  Higher values for $\alpha^{(n)}$
seem slightly favoured~\cite{GRSb}.

The gap between our extraction of the polarisabilities and the $\chi$PT
results is well known, see e.g.~\cite{KM99}. It can be traced back to the fact
that the Compton scattering data at large angles is systematically more and
more enhanced over forward scattering as the photon energy is
increased~\cite{Lucas,Hornidge:2000xs}. This favours larger values of
$\beta_0$.



\absatz Although the scarcity of data at low energies is a big hindrance, the
comparison of the present calculation to experiment shows that it is quite
appropriate to determine neutron scalar polarisabilities from low energy
Compton scattering. Our approach uses EFT and hence it is model-independent
and allows one to estimate the theoretical accuracy as $\sim 20\%$ for the
polarisabilities, assuming higher order contributions are of natural size. 
Future higher precision experiments e.g.~at TUNL in the
energy r\'egime $\omega\sim 25-50\;\MeV$ would be most useful. As one can see
from Fig.~\ref{comptonfit.fig}, at lower photon energies, the polarisability
contributions become smaller than \NtwoLO, while higher energies introduce
large theoretical errors since corrections going like $\omega/\Lambda$ are not
suppressed sufficiently strong any more. The energy r\'egime proposed is hence
an interesting window of opportunity to determine nucleon polarisabilities in
a model-independent way without having to deal with pions as explicit degrees
of freedom. Small angle scattering data constrains $\alpha_0+\beta_0$ and
hence can serve as a cross check to the Baldin sum rule, whereas
backscattering data probes $\alpha_0-\beta_0$ and hence is needed for a better
determination of $\beta_0$. In a future publication~\cite{GRSb}, we will
address details of the calculation as well as a more accurate estimate of
N${}^3$LO effects.  The only undetermined operators entering at N${}^3$LO are
$(C_E \vec{E}^2+C_M\vec{B}^2) (N^T P_iN)^\dagger (N^T P_iN)$ with projections
$P_i$ onto the $^3S_1$ channel.  These operators contribute to residual
electric and magnetic deuteron polarisabilities, i.e.~to those not generated
by polarisation of the two nucleons against each other or by polarising the
nucleons themselves.


\section*{Acknowledgements}

We are much indebted to discussions with the EFT group at the Institute for
Nuclear Theory and the Physics Department at the University of Washington in
Seattle, particularly with M.~J.~Savage, as well as with Th.~Hemmert and
N.~Kaiser. H.W.G.~wants to emphasise especially the hospitality of the INT
during the Summer 2000 Workshop on ``Effective Field Theories and Effective
Interactions'', and its financial support. The participants of the workshop
created a highly stimulating atmosphere. We also acknowledge support in part
by the Bundesministerium f{\"u}r Bildung und Forschung (H.W.G.), the DFG
Sachbeihilfe GR 1887/1-1 (H.W.G.), the US Department of Energy grant
DE-FG03-97ER41014 (G.R.) and the Natural Sciences and Engineering Research
Council of Canada (G.R.).



\end{fmffile}
\end{document}